\pdfoutput=1
\pdfmapfile{+SourceSerifPro.map}
\documentclass{adonis}
\usepackage{natbib} 
\usepackage{lettrine}
\AtBeginEnvironment{quote}{\par\small} 

\title{A Bureaucratic Theory of Statistics}
\author{Benjamin Recht}
\affiliation{University of California, Berkeley}
\correspondence{brecht@berkeley.edu}
\version{\today}

\runningauthor{Recht}
\runningtitle{A Bureaucratic Theory of Statistics}

\abstract{This commentary proposes a framework for understanding the role of statistics in policy-making, regulation, and bureaucratic systems. I introduce the concept of ``ex ante policy,'' describing statistical rules and procedures designed before data collection to govern future actions. Through examining examples, particularly clinical trials, I explore how ex ante policy serves as a calculus of bureaucracy, providing numerical foundations for governance through clear, transparent rules. The ex ante frame obviates heated debates about inferential interpretations of probability and statistical tests, p-values, and rituals. I conclude by calling for a deeper appreciation of statistics' bureaucratic function and suggesting new directions for research in policy-oriented statistical methodology.}

\begin{document}

\maketitle

\vspace{-16px}
\noindent{\small {\bf keywords:} {ex ante policy, statistical inference, statistical testing, bureaucracy.}}

\vspace{32px}

\lettrine[lines=3,nindent=0em]{A}{s a discipline,} Statistics aspires to many lofty goals. Adherents of Judea Pearl or Don Rubin might tell you that Statistics provides a science of causation. Yoav \citet{Benjamini16} argues that Statistics ``offers a first line of defense against being fooled by randomness, separating signal from noise.'' Philip \citet{Stark2024} frames Statistics as the field of computational epistemology, an algorithmic system for divining what is likely to be true.

Despite a century of rigorous development of statistical hypothesis testing and causal inference methods, it is not at all clear that Statistics actually does any of these things. We see something different when we look at the practice of statistical testing and statistical inference. Cyberneticist Stafford \citet{Beer79} famously asserted that “the purpose of a system is what it does.” If, after 100 years of fighting and browbeating, we see that statistical testing consistently fails to provide strong epistemological guarantees or determinations of causation, then it’s counterproductive to continue teaching our students that this is the purpose of statistics. But then, what exactly do statistical tests do?

To get a sense of how to answer this question, let me focus on the crown jewel of Statistics---what some statisticians call  “the gold standard of causal inference”---the randomized control trial (RCT). How are RCTs used in practice? The biggest success of the RCT has been in drug trials, providing evidentiary guidelines for pharmaceutical companies to demonstrate that their drugs are safe and effective. Randomized clinical trials have been more generally applied in medicine to establish guidelines for standard of care. More recently, technology companies have widely adopted RCTs to evaluate whether to deploy new software features. And controversially, statistical tests have transformed the publication of scientific literature, providing folk standards for peer review.

In all of these cases, causation is a secondary concern. The RCT primarily serves as a mechanism of regulation, regulating which drugs come to market, which clinical practices become standard of care, which software goes into production, and which academic papers prestige journals publish. That is, one of the prime roles of the RCT is facilitating rulemaking. 

Rulemaking is certainly not the singular valuable application of statistics, but it has been a revolutionary and uncelebrated use. Causal inference researchers, whether potential outcomes advocates for whom all causation stems from hypothetical RCTs, or do-intervention scholars who still argue RCTs should have less relevance, should reckon with this underappreciated role of statistical theory.

Statistics sells itself short by not embracing this role. From its inception, statistics has been the mathematics of bureaucracy. It provides a numerical foundation for governance by clear, transparent, aggregate rules. Statistics helps governments measure what experts on the ground see and create reasonable metrics for consensus to move forward with policy.

In this commentary, I introduce my own umbrella term for such statistical rulemaking: \emph{ex ante policy}. Ex ante because the rules and procedures are designed before data collection. Policy because the rules aim to inform actions to be taken after data collection. As I’ll describe, most common Frequentist methods, including testing and confidence intervals, are examples of ex ante policy. Many Bayesian methods qualify as well. Instead of belaboring the weary Frequentist/Bayesian divide, I sharply contrast ex ante policy against ex post inference, where we draw conclusions about theories or parameters after observation. Highlighting this ex ante/ex post distinction not only clarifies the theoretical guarantees of statistical methods but also decouples scientific uncertainty from decision making. The framing of statistical ex ante policy opens many potential research directions for statisticians in the human-facing sciences.

\section{Ex ante policy}

By ex ante policy, I mean the specification of rules by some regulatory body that governs acceptable future actions. The example likely most familiar to statisticians is the evidentiary standards of the FDA for drug approvals in the United States. The 1962 Kefauver-Harris Amendment to the Federal Food, Drug, and Cosmetic Act demands \emph{substantial evidence} of efficacy for the approval of a new drug. Though it is nowhere explicitly written, the FDA and pharmaceutical companies have interpreted this evidentiary standard as two randomized controlled clinical trials with results significant at the 0.05 level~\citep{kennedy-shaffer_when_2017}.

The timely influence of A. Bradford Hill (1897-1991) coincided with the FDA’s need to create regulatory standards after the Federal Food, Drug, and Cosmetic Act of 1938. Not only did Hill design the first randomized clinical trial in 1946, but he consulted with global health leaders, including American institutions like the FDA and National Cancer Institute, about reasonable guidelines for trial execution and interpretation. His consultation was pivotal in the implementation and acceptance of randomization as a central part of the mammoth Salk Vaccine trial.

Hill explicitly viewed randomized trials as a means of informing policy. He emphasizes design and randomization as tools for bias mitigation when evaluating best medical practice or pharmaceutical efficacy. Perhaps surprisingly, in his writing aimed at the medical community, Hill almost never talks about p-values. The p-values were to convince statisticians. Let me break from his tradition here, as I want to unpack something valuable about the Fisherian application of p-values in RCTs.

In a randomized trial, we want to determine whether we want to use a treatment. We declare in advance that we will recommend the treatment if it clears an evidentiary bar of efficacy. We agree upon some metric, gather $n$ individuals, and assign them at random to treatment.

Let $Y_i$ denote the measured outcome of each individual and $X_i$ their treatment assignment. Let $R(X,Y)$ denote the Pearson r-coefficient between the $n$-vectors $X$ and $Y$. An ex ante policy would be to agree to approve the use of the treatment if 
\begin{equation}\label{eq:correlation-threshold}
R(X,Y)  \geq \frac{t}{\sqrt{n}}
\end{equation}
for a specified threshold $t$. If the Y are binary outcomes and we set $t = 1.96$, this acceptance criterion is the same as the proportions z-test at the level $\alpha=0.05$. It is also equivalent to the significance threshold of the chi-squared test. A minor modification in the formula for the threshold gives a t-test or a z-test with nonbinary treatments and outcomes.\footnote{For example, if $X$ and $Y$ are not binary outcomes, then the z-statistic can be defined in terms of the coefficient of regression of $Y$ on $X$, $\hat{\beta}$.In this case, the z-statistic is $\hat{\beta}/\text{SE}(\hat{\beta})$, and the reformulation of equation~\eqref{eq:correlation-threshold} becomes $R(X,Y)  \geq t/\sqrt{n + t^2-2}$.}  Such formulas reframing tests in terms of correlation can be found in Cohen’s \emph{Statistical Power Analysis for the Behavioral Sciences} [1969].\nocite{CohenBook}

% [footnote, the formula for the threshold at level alpha is t >= t_{alpha,n}/\sqrt{t_{alpha,n-2}^2/n + 1}]

Regardless of how we set the threshold, the policy states---in advance---that if the correlation between the treatment assignment and the outcome is large enough, the treatment is to be declared efficacious.

The explanation for why this is a reasonable policy isn’t particularly deep. For any outcome, the expected magnitude of the correlation coefficient with a random treatment assignment is $n^{-1/2}$. Observing a higher correlation would be unlikely if the treatment truly had no effect.

Statisticians have bickered about the right value of the threshold $t$ for the right-hand side of Equation~\eqref{eq:correlation-threshold}. We all agree the threshold should be larger than 1. Computing a precise value to multiple significant figures is fine for publication in the \emph{Annals of Statistics}, but likely an unnecessary paralysis by analysis for most applications. Whether you use a Fisher Exact test or a chi-squared test, you are quibbling over an evidentiary bar. We have to pick a convention. The threshold might as well be 2. The z-test and chi-squared test are convenient as they are trivial to compute. And you can guarantee it will be an accurate approximation by demanding as part of the rules that n is sufficiently large. Power analysis is, of course, also ex ante policy.

The value of such regulatory rules goes well beyond the associated statistical guarantees. As Bradford Hill repeatedly noted, randomization removes potential biases and confounders in the trial. Time of randomization sets clear start points for survival analyses. 

Moreover, there are social considerations in policy making that may be met by RCTs. Ex ante guarantees provide transparency. And through this transparency, they bring some notion of fairness. Aleatoric randomness has been a powerful tool for fair decision-making for thousands of years. Assignment by a lottery is fair in the narrow utilitarian sense that everyone is treated indistinguishably and no party can game the system. Parties that agree to such random allocation, whether they be sports teams flipping coins to decide who acts first or patients hoping to be assigned to experimental treatment, agree to these rules even though there is a chance they will ex post receive a suboptimal assignment.  What we mean by fairness is a moral judgment that must be deliberated upon by all affected parties. But transparent rules let us adjudicate endorsed notions of fairness and ensure accountability.

Ex ante statistical guarantees additionally assure stakeholders that---at least most of the time---our study should produce a trustworthy answer. Accepting possible error is part of policy making. We know there will always be some uncontrollable epistemic uncertainty. Adding aleatoric uncertainty to mitigate the epistemic uncertainty of confounding is a reasonable tradeoff.

All of the preceding arguments about p-values apply to confidence intervals as well. The definition of a confidence interval is always a confusing mouthful: A confidence interval is a set computed from empirical data that is ex ante guaranteed to contain a parameter of interest with some probability. This probability is defined over the sample space of potential future outcomes. 

I prefer to define confidence intervals as randomized algorithms. A confidence interval is a clearly specified procedure that combines generated randomness, interaction with the world, and numerical algorithms to yield a set. The algorithm succeeds if the set contains the parameter we care about and fails otherwise. We ex ante characterize the probability of success in terms of the prospective randomness. Despite their persistent misinterpretation as otherwise (see, for example,~\citet{hoekstra_robust_2014}), confidence intervals \emph{only} provide ex ante guarantees. We have no means of checking if the confidence interval algorithm succeeded, but we can guarantee that success is likely.

My definition of ex ante policy is not restricted to frequentist tests. Bayesian decision theory is also built upon ex ante guarantees. A proper Bayesian must specify a coherent prior, statistical models of outcomes, and a utility function. With these ingredients defined before action, the optimal decision can be computed by maximizing expected value.

For example, we can set a medical policy of giving someone a drug using a Bayesian framework. The prior could be the prevalence of a disease, the likelihood functions could be the sensitivity and specificity of the test, and the utilities could weigh the costs and benefits of treating the healthy and not treating the sick. Put together, a likelihood ratio test, specified in advance, determines which patients receive treatment.

Though often given too much credence as a general approach to decision making, the rational choice framework buys you a path forward for policy. If a group of stakeholders agrees that a rational policy is acceptable, then this gives way to a computational solution, whether that policy dictate rules for medical residency matching or organ transplantation.  Ex ante rational decision making may lead to bad outcomes, but as was the case with random assignment, the actions are chosen in a specified, principled manner before the results are known. Some people can assuage themselves knowing they made the right decision no matter the outcome. I won’t dwell on whether or not this is virtuous. It might not be perfect, but if you agree upon ``rational,'' you at least establish a potential solution. 

A key feature of ex ante policy is that it forces policymakers to state their models, their means of intervention, and what they value. This not only lets us plan under reasonable models of future uncertainty, but also lays out rules for informed debate. If some members of a community want stronger evidentiary standards, they can propose their own ex ante rules and lobby for their adoption. A community of stakeholders must decide together which system of ex ante policy, be it frequentist tests, rational choice, or some other means of participatory decision making, best aligns with their values.

\section{The ex ante/ex post confusion}

Statistics asks many different kinds of questions, but we confuse our students because the methods often look the same. Are we trying to quantify the truth of a hypothesis? Or are we trying to quantify the error in a measurement to aid decision making? We need to speak with clarity about this. I don’t want to blame anyone, but until we disambiguate use, we will have endless arguments about what p-value thresholds mean for the replicability of science.

Isolating ex ante policy from other tasks alleviates many of the most confusing points of statistics. We should make a clear distinction between ex ante policy and ex post inference. Ex post inference, drawing conclusions about the verisimilitude of theory or the nature of material reality from empirical evidence, is a task distinct from ex ante policy. Ex post inferences could be Popperian falsification, a Bayesian update of posteriors, or the data-driven inference that ``A causes B.'' Whereas ex ante policy comes with verifiable theorems, ex post inference does not. Ex post inference is complex and subjective. Ex ante policy is prescriptive and objective.

Highlighting ex ante policy helps assuage another common critique of statistical tests: that deployed statistical procedures are ritualistic. \citet{GIGERENZER2004587} argues that the practices in his field of psychology are a confusing mishmash of contradictory ideas from Fisherian testing and Neyman-Pearson decision making. He calls the null hypothesis significance test, ubiquitous in experimental psychology, the “null ritual” and argues how it’s not based upon sound statistical theory. Regardless of whether the null ritual has an epistemological grounding, it undeniably provides a common rulebook for assessing submitted papers. By changing the word we use to describe practice, we can get a more productive perspective. Instead of calling them rituals, we should refer to RCTs and their associated tests as \emph{regulations}. We are better off articulating the rules rather than pretending they don’t exist.

Gigerenzer also argues that the null ritual is ill-suited as a foundational algorithm for the scientific method. There is ample evidence on his side, or at least not substantial evidence to the contrary. While statistics has admirable aspirations to help answer questions about ex post inference, it’s hard to find grand scientific discoveries solely enabled by RCTs or other causal inference methods. Scientific inference is rarely numerical and always cultural and heuristic. Not only is there no universally accepted method of severe testing for Popperian falsification, but science doesn’t progress via Popperian means. Moreover, there are no computable Bayesian updates that crunch empirical data into crisp posteriors. The probabilities of scientific theories live in our heads, and how we update our credence for them doesn’t appear to have a clean numerical algorithm. This is fine. Scientists and investigators can do amazing things without being computers. Policy may necessitate inflexible rules, but scientific inquiry does not.

\section{What do the experts say?}

I’m not sure anything I’ve written so far is particularly controversial. In fact, it’s not hard to find arguments about necessary conditions for policymaking throughout the causal inference literature. For example, in the first paragraph of the introduction to Causal Inference for Statistics, Social, and Biomedical Sciences, \citet{Imbens2015Causal} assert their focus will be on questions of policy:
\begin{quote}
... a medical researcher may wish to find out whether a new drug is effective against a disease. An economist may be interested in uncovering the effects of a job-training program on an individual’s employment prospects, or the effects of a new tax or regulation on economic activity. A sociologist may be concerned about the effect of divorce on children’s subsequent education.
\end{quote}
Though Pearl advocates for a different approach to causal inference, he writes with Mackenzie in \emph{The Book of Why} that his primary interest in studying causal inference was to answer thorny policy questions. ``[As] we began posing causal questions in complex legal, business, medical, and policy-making situations, we found ourselves lacking the tools and principles that mature science should provide.'' 

No matter where you look, the answer to a fundamental why question, ``Why are we invested in methods of causal inference?'' is ``Because we want to help policy-makers.''

\citet{Rubin2022} emphasizes his policy-centric views in a recent interview with the journal \emph{Observational Studies}:
\begin{quote}
In order to think about policy recommendations, I believe you have to be realistic and try to figure out the range of things you might recommend as a consequence of your analysis. Then, you construct the observational study and associated hypothetical randomized experiment so that these studies address the intervention that you might recommend, the interventions that you could actually implement. This is needed if you are going to get meaningful confidence interval estimates or p-values. Otherwise they are all model-based inferences. Where did the model come from? As George Box said ``All models are wrong, but some are useful.'' And earlier as John von Neumann said ``Truth is much too complicated to allow anything but approximations.''

For any analysis, you are telling a story. If you want frequentist p-values and confidence intervals from an observational study, you have to tell me a story about the hypothetical randomized experiment… You have to tell me a story that makes a plausible and relevant case for its relevance.
\end{quote}

Rubin is arguing my case in this passage. Statistical arguments can all be story-telling and correlation, but such stories must follow appropriate rules. If we talk about why RCTs are suitable for policy, we can focus on what it would mean for observational studies to create those policies.
\section{Toward a Bureaucratic Theory of Statistics}

Framing statistical testing as policymaking leads to many questions for methodologists of statistics and causal inference. What statistical rules should we advocate for? Do current designs, whether they be panels, regression discontinuities, instrumental variables, or differences and differences, suffice for policymaking? Can we design rules in advance that might make such methods appropriate? Is it ever acceptable to set policy based on observational data? A systematic articulation of ethical issues and ex ante guarantees might be preferable to the more ad hoc ways policies are currently determined. Observational or modeling studies are certainly used to inform guidelines (for example, the United States  Preventive Services Taskforce routinely uses modeling simulations based on observational data), and the question remains: should we use them that way? 

If we can agree that the primary goal of what we have been calling “causal inference” is the determination of policy, we can proceed to ask what makes our methods best suited to policy evaluation and what makes our methods most suited to being rules in an elaborate bureaucratic game. Recent research along the lines of~\citet{Banerjee2020} points to an abundance of fruitful methods work to pursue. In particular, an important property of policy and rules is that they can be changed. If a particular suite of methods leads to outcomes misaligned with our values, we can adapt them.

In a 2021 editorial, Philosopher Deborah Mayo\nocite{Mayo2021}, who favors a falsificationist view of frequentist testing, argues, “The key function of statistical tests is to constrain the human tendency to selectively favor views they believe in.” I would put it only slightly differently.  Statistical tests constrain outcomes in participatory systems. Engineers want to push features to get promoted; data science teams insist on AB tests to ensure these features don’t harm key metrics. Drug companies want to make a ton of money; clinical trials ensure drugs aren’t harmful and have a chance of being beneficial. Academics want to publish as many papers as possible to get their h-index to the moon; journals insist on some NHSTs to placate editors. The purpose of statistical tests is regulation.

“Scientific” questions about people, whether in medicine, political science, economics, or sociology, concern questions about what we ought to do as a society.  Social science and medicine are not interested in discovering causation. They are looking for evidence to justify policies that will impact populations of people. And statistics has been one of the most powerful methods for population-level, bureaucratic assessment.

No one wants to be called a bureaucrat. It takes on a disparaging connotation, especially in academia, where besieged professors are downtrodden by exponentially growing paperwork. And yet, bureaucracies enable massive systems of governance to function. These systems do not always function well, but they operate at astounding scales. It’s hard to take pride in bureaucracies, but what if we embraced the admirable goal of creating well-run systems of participatory decision making at a global scale?

\section*{Acknowledgements}

BR would like to thank Avi Feller, Chris Harshaw, Jessica Hullman, Tengyuan Liang, Deborah Mayo, and Nati Srebro for helpful conversations that inspired these arguments.

\bibliography{bureaucratic}
\bibliographystyle{plainnat}
\end{document}